\documentclass[12pt,a4paper]{article}
\usepackage{graphicx,color}
\usepackage{amsmath}
\usepackage{amssymb}
\pdfoutput=1
\usepackage{multirow}
\usepackage{url}
\usepackage{subfigure}
\usepackage{authblk}
\usepackage{fullpage}
\usepackage{lscape}

\begin{document}
\vspace*{-0.5in}
\thispagestyle{empty}

\vspace{0.5in}
\begin{center}
\begin{LARGE}
{\bf Status Report:\\
A Search for Sterile Neutrino at J-PARC MLF (E56, JSNS$^2$)\\}
\vspace{5mm}
\end{LARGE}
\begin{large}
July 8, 2015\\
\end{large}
\vspace{5mm}
{\large
M.~Harada, S.~Hasegawa, Y.~Kasugai, S.~Meigo,  K.~Sakai, \\
S.~Sakamoto, K.~Suzuya \\
{\it JAEA, Tokai, JAPAN}\\
\vspace{2.8mm}
E.~Iwai, T.~Maruyama\footnote{{Spokesperson: 
(takasumi.maruyama@kek.jp) }}, 
S.~Monjushiro, K.~Nishikawa, M.~Taira\\
{\it KEK, Tsukuba, JAPAN}\\
\vspace{2.8mm}
M.~Niiyama\\
{\it Department of Physics, Kyoto University, JAPAN}\\
\vspace{2.8mm}
S.~Ajimura, T.~Hiraiwa, T.~Nakano, M.~Nomachi, T.~Shima\\
{\it RCNP, Osaka University, JAPAN}\\
\vspace{2.8mm}
T.~J.~C.~Bezerra, E.~Chauveau, H.~Furuta, F.~Suekane\\
{\it Research Center for Neutrino Science, Tohoku University, JAPAN}\\
\vspace{2.8mm}
I.~Stancu\\
{\it University of Alabama, Tuscaloosa, AL 35487, USA}\\
\vspace{2.8mm}
M.~Yeh\\
{\it Brookhaven National Laboratory, Upton, NY 11973-5000, USA}\\
\vspace{2.8mm}
H.~Ray\\
{\it University of Florida, Gainesville, FL 32611, USA}\\
\vspace{2.8mm}
G.~T.~Garvey, C.~Mauger, W.~C.~Louis, G.~B.~Mills, R.~Van~de~Water\\
{\it Los Alamos National Laboratory, Los Alamos, NM 87545, USA}\\
\vspace{2.8mm}
J.~Spitz\\
{\it University of Michigan, Ann Arbor, MI 48109, USA}
}
\end{center}
\renewcommand{\baselinestretch}{2}
\large
\normalsize

\setlength{\baselineskip}{5mm}
\setlength{\intextsep}{5mm}

\renewcommand{\arraystretch}{0.5}

\newpage

\tableofcontents
\vspace*{0.5in}
\setcounter{figure}{0}
\setcounter{table}{0}
\indent

\section{Introduction}
\indent

On April 2015, the J-PARC E56 (JSNS$^2$: J-PARC Sterile Neutrino Search using 
neutrinos from J-PARC Spallation Neutron Source) experiment officially 
obtained stage-1 approval 
from J-PARC. We have since started to perform liquid scintillator R$\&$D 
for improving energy resolution and fast neutron rejection. Also, we are 
studying Avalanche Photo-Diodes (SiPM) inside the
liquid scintillator. In addition to the R$\&$D work, a background 
measurement for the proton beam bunch timing using a small 
liquid scintillator volume was planned, and the safety discussions for 
the measurement have been done. This report describes the status of 
the R$\&$D work and the background measurements, in addition to the 
milestones required before stage-2 approval. 

\section{\setlength{\baselineskip}{4mm}
Experimental milestones before stage-2 approval}
\indent

There are two main points to show the J-PARC PAC before stage-2 approval.\\
(1) The concrete detector location should be considered including the physics 
potential of the experiment. This was written in the minutes of the 19th
PAC~\cite{CITE:PAC19report}. 
For this purpose, we have to discuss this issue in close coordination
with the facility people, since the constraints from the facility are
important to consider alongside the physics ones.\\
(2) We assume a rejection factor of 100 for fast neutron
events induced 
by cosmic rays. The technique using Cherenkov light at LSND
already achieved this factor, however we have to show the performance of the
liquid scintillator (and photo-detector) quantitatively. Both the Cherenkov 
technique and the Pulse Shape Discrimination (PSD) technique are tested for this purpose.
\\ 
 
For the former, we would like to have a special joint discussion with PAC members
and MLF facility people (the FIFC is a candidate). The discussion includes 
safety issues as well as facility maintenance. 
We will show the status of the R$\&$D for the latter.

\section{R$\&$D for the JSNS$^2$ Detector}

\subsection{Motivation}
\indent

As shown in the status report submitted to the PAC last 
November~\cite{CITE:NOVSR, CITE:ARXIV, CITE:PTEP}, all backgrounds 
in the candidate detector location are manageable for performing the 
JSNS$^2$ experiment. One crucial assumption is that 
the detector should have a rejection factor of 100 
for fast neutron events induced by cosmic rays.  
Table~\ref{TAB:grandsum} shows a summary of the number of
background and signal events in the JSNS$^2$ (shown 
in reference~\cite{CITE:NOVSR}). 
Based on these assumptions, the background rate induced by cosmic rays
is comparable to the other background rates.

\begin{table}[htbp]
\begin{center}
\begin{tabular}{|c|c|c|}\hline
&Contents& events\\ \hline \hline
&$sin^22\theta=3.0\times10^{-3}$&\\
&$\Delta m^2$=$2.5eV^{2}$ & 480\\ 
Signal&(Best fit values of MLF)&\\ \cline{2-3} 
&$sin^22\theta=3.0\times10^{-3}$&\\
&$\Delta m^2=1.2eV^{2}$&342\\
&(Best fit values of LSND)&\\\hline\hline
&$\overline{\nu}_{e}$ from $\mu^{-}$&237\\\
&$^{12}C(\nu_{e},e^{-})^{12}N_{g.s.}$&16\\\
background&beam-associated fast n&$\le$13\\
&Cosmic-induced fast n&37\\
& Accidental coincidence &32\\\hline
\end{tabular}
\caption{\setlength{\baselineskip}{4mm}
Number of events in 50~tons.
1MW beam $\times$ 5 years is assumed.
}
\label{TAB:grandsum}
\end{center}
\end{table}

The factor of 100 fast neutron rejection capability was already achieved by the LSND experiment 
using Cherenkov light detection, since protons in the energy range of 20-60 MeV (recoiled by fast neutrons induced by cosmic rays) cannot emit Cherenkov light. 
However we have to show the performance of JSNS$^2$ 
detector independently since the design is different from the LSND. 

We are also trying to establish a PSD technique, because it uses independent
information from Cherenkov light: The output signals' pulse shape 
from liquid scintillator are different between protons recoiled by 
fast neutrons and positrons from Inverse Beta Decay (IBD). 
If we can use both information from Cherenkov and PSD, the rejection power could be even stronger
than what has been assumed. 

In the following subsections, we show the status of the R$\&$D using
small prototype devices. For the Cherenkov test, there is a $\sim$10L prototype,
and a 0.1L level prototype is used for the PSD test.  

The possibility to use APD (SiPM) is also being considered, since
the space of the detector is tightly restricted, thus SiPMs may be 
used for the veto region even though the photo-coverage is small. 
Therefore, its R$\&$D status is also described briefly.

\subsection{PID Capability using Cherenkov Light in Liquid Scintillator}
\label{SEC:SC}
\indent

\begin{figure}[htbp]
\centering
\subfigure[Setting for Cherenkov + scintillation]{
\includegraphics[width=0.45\textwidth,angle=0]{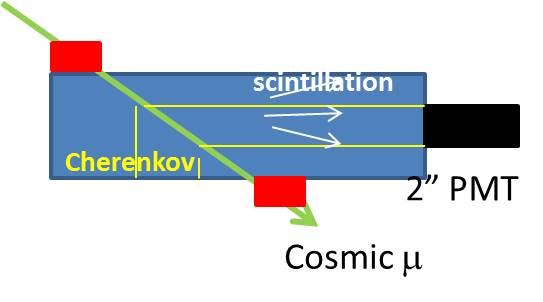}
}
\subfigure[Setting for scintillation light only]{
\includegraphics[width=0.44\textwidth,angle=0]{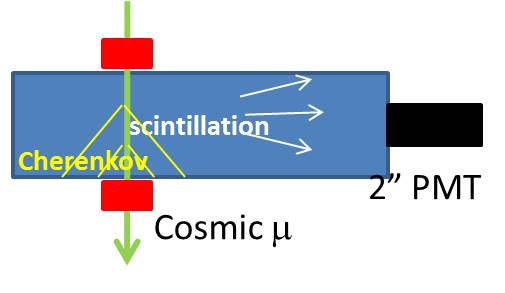}
}
\caption{\setlength{\baselineskip}{4mm}
Schematics of the test setup. (a) shows the setting for the Cherenkov
plus scintillation detection and (b) shows the setting for the scintillation 
light only detection. The cylindrical prototype with 
a size of 130mm (diameter) $\times$ 1000mm (height) is filled with water 
or diluted scintillator (blue parts of the schematics).
Scintillation counters, which are shown as red boxes, 
are used to tag cosmic ray muons. Coincidence signals by the scintillation 
counters are used as the reference timing for the Cherenkov and scintillation 
light inside the cylinder.
The effective area for each scintillator
is 5 $\times$ 10 cm$^2$. The average muon path distance from the 2 inch PMT
was kept as 65 cm to make the amount of scintillation light the same in both settings.
} \
\label{FIG:SETUPCS}
\end{figure}

The JSNS$^2$ experiment plans to use linear alkylbenzene (LAB) as the 
base of the liquid scintillator. However,  
we have started to follow the measurements in reference~\cite{NIM_LSND} 
using Mineral oil + 0.03g/l b-PBD at first, since
the LSND study will be an excellent reference for subsequent measurements.
A cylindrical prototype detector with dimensions of 130mm (diameter) $\times$ 
1000mm (height) was prepared for this test. 
Figure~\ref{FIG:SETUPCS} shows the measurement setup.
The cylindrical prototype 
is filled with water or diluted scintillator, and a black
sheet is lining inside the detector to avoid reflections of light.
The motivation of this experiment is to measure the timing difference 
between Cherenkov light and scintillation light mainly. The Cherenkov light 
has a faster emission timing than that of the scintillation light in general. 
Note that water does not emit scintillation light, therefore it provides a good 
cross check of the timing information.

If the photo-detector receives the scintillation and Cherenkov light at the
same time, the timing distribution should be like Fig.~\ref{Fig:timing}.  
The solid line shows the scintillation light while the dashed line corresponds to 
the Cherenkov light.
Here we assume that the time distribution of the scintillation light emission 
is the sum of a exponential for the fast time component
and a (1+t/$\tau$)$^{-2}$ empirical shape of the slow light, 
as written in the LSND NIM paper~\cite{NIM_LSND}\footnote{\setlength{\baselineskip}{4mm} The fast time constant is 1.65 ns, and the slow time constant is 
22.58 ns. The mixture ratio is 57$\%$ and 43$\%$ for each component.}.
We also assume a time resolution of 0.5ns for the photo-detector 
(only for Cherenkov). Note that these time constants depend on the liquid scintillator being used.

\begin{figure}[h]
 \centering
 \includegraphics[width=0.45 \textwidth]{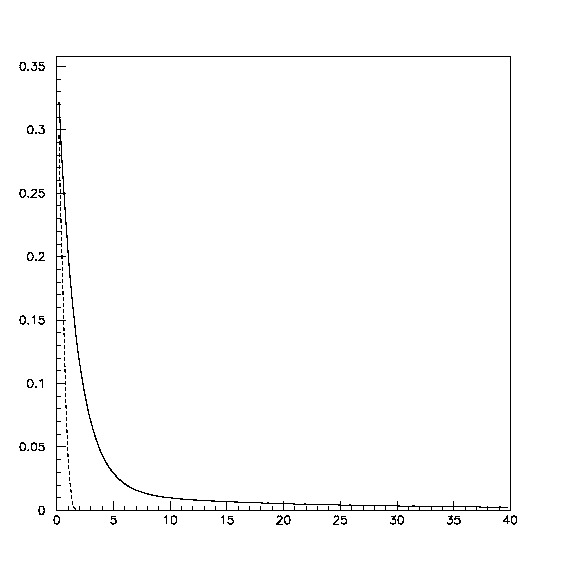}
\caption{\setlength{\baselineskip}{4mm}
Timing distributions of the scintillation light (solid) and Cherenkov light (dashed). The horizontal axis corresponds to the time, unit is ns.
}
 \label{Fig:timing}
\end{figure}

\subsubsection{Cross-check with filling Water}
\indent

For the water case, there is no scintillation light, and we therefore did not see 
any light in the setup shown in the bottom of Fig.~\ref{FIG:SETUPCS}(b).
On the other hand, we can see Cherenkov light with the setup shown
in the top of Fig.~\ref{FIG:SETUPCS}(b).

\begin{figure}[htpb]
 \centering
 \includegraphics[width=0.55 \textwidth]{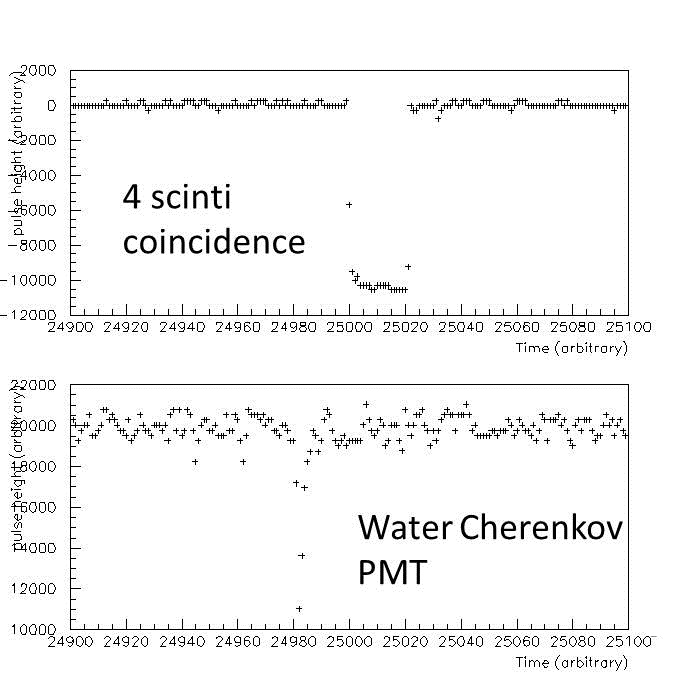}
\caption{\setlength{\baselineskip}{4mm}
Typical timing distributions of the coincidence from the scintillation counters (top) and Cherenkov light (bottom) measured by the oscilloscope. The horizontal axis corresponds to the time. One count is 2 ns. The vertical axis of the plots is the pulse height counts. 1000 counts corresponds to 1mV in the bottom plot for this case.
}
 \label{Fig:raw_timing}
\end{figure}

Figure~\ref{Fig:raw_timing} shows the typical waveform timing distributions in an
oscilloscope. We will use the oscilloscope data for the analysis since
the cosmic ray trigger rate is very low.
The two plots show typical timing distributions of the coincidence from 
the scintillation counters (top) and Cherenkov light (bottom). 
Note that the coincidence signal from the scintillation counters was 
delayed due to electronics. Also, this relative timing was smeared by the 
PMTs jitter ($\sim$1 ns), electronics (within 1 ns), and hit position of the
cosmic rays ($<<$ ns). This measurement is important for understanding the
effect of the jitters.    

If the pulse height exceeds below -2000 counts beyond the 
ground line (e.g.: $\sim$20000 is the ground line, and the threshold is
$\sim$18000 in the bottom plot of Fig.~\ref{Fig:raw_timing}),
it is regarded as signal.
-2000 corresponds to a threshold
of about 1/3 photo-electrons (p.e.), therefore the threshold is low enough to 
search for the signal hits.

Figure~\ref{Fig:wc_timing} shows the relative timing of water Cherenkov signals
with respect to the coincidence. 
The timing of the fastest point 
of each signal which exceeds 2000 counts is plotted in this figure.
Out of 100 triggers, about half of them have signals beyond the 
threshold, and this rate is reasonable according to the MC simulation\footnote{\setlength{\baselineskip}{4mm}
MC calculation shows that the Cherenkov light gives 0.5$\sim$0.8 p.e. / trigger.
 The uncertainty is due to the momentum and angle of cosmic rays. They are not 
included in the MC yet because the timing does not depend on those.}.
As shown in the plot, the total jitter of the water Cherenkov signal 
(or the relative timing resolution from systematic effects) is $\sim$ 1.8 ns. 

\begin{figure}[htpb]
 \centering
 \includegraphics[width=0.45 \textwidth]{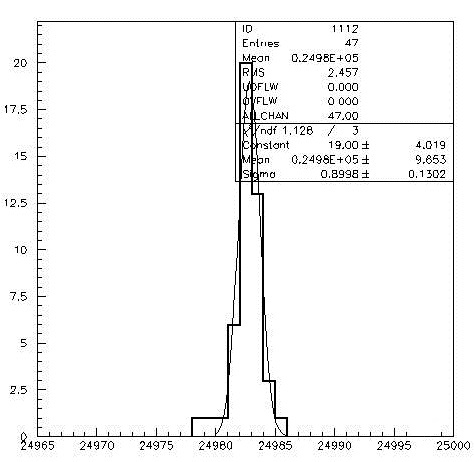}
\caption{\setlength{\baselineskip}{4mm}
Timing of the Cherenkov light. The horizontal axis is time counts, and 1 count corresponds to 2 ns. With respect to the coincidence signal, the total jitter is $<$ 0.9 counts (1.8 ns) from the fit result as shown in the bottom of the fit parameters.  (0.8998$\pm$0.1302)
}
 \label{Fig:wc_timing}
\end{figure}

\subsubsection{Diluted Scintillator}
\indent

Hereafter we show the measurements using a diluted scintillator, which contains
mineral oil plus 0.03g/l of b-PBD. This diluted scintillator was used in the 
LSND experiment, and it can be used as the reference of the other measurements
or for discussion of detector performance.

\subsubsection*{\setlength{\baselineskip}{4mm}
Setting for scintillation light only detection (Fig.~\ref{FIG:SETUPCS}(b))}
\indent

\begin{figure}[htbp]
\centering
\subfigure[Typical event 1 (slow timing)]{
\includegraphics[width=0.45\textwidth,angle=0]{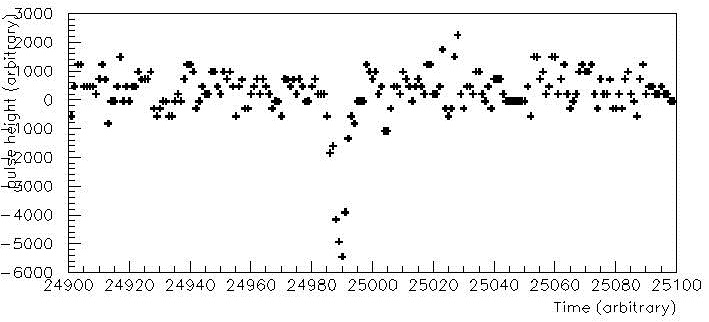}
}
\subfigure[Typical event 2 (multiple light)]{
\includegraphics[width=0.45\textwidth,angle=0]{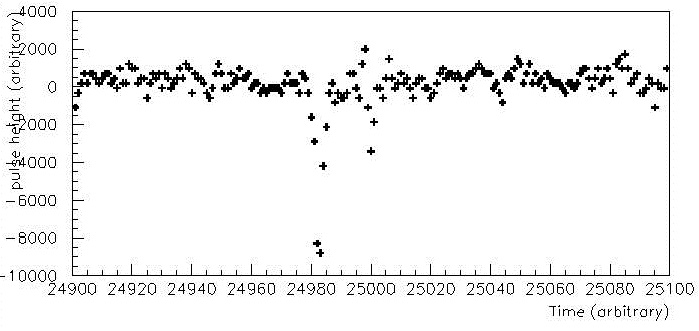}
}
\caption{\setlength{\baselineskip}{4mm}
Typical signals using diluted scintillator. 
The horizontal axis is time counts, and 1 count corresponds to 2 ns.
(a) We see a slow timing hit,
and (b) multiple hits are seen. Note that the Cherenkov signal is seen around 
$\sim$24982 counts on the horizontal axis as shown in Fig.~\ref{Fig:wc_timing} 
} 
\label{FIG:SONLY}
\end{figure}

The scintillation light with the setting shown in 
Fig.~\ref{FIG:SETUPCS}(b) is discussed first. With this setting, 
there should be only scintillation light detection. 
Figure~\ref{FIG:SONLY} shows the typical signals using the diluted scintillator.
Events with a slow scintillator timing signal and multiple light signal 
are shown.

To analyse events with multiple hits, one hit is defined as follows.
\begin{itemize}
\item Points less than -2000 counts are searched for.
\item If there are more than two points, it is judged whether they are
     in sequence in time or not. If they are in sequence, they are regarded
     as one hit.
\item The fastest point within one hit is recorded as the ``timing of the 
     hit''   
\end{itemize}  

\begin{figure}[htbp]
\centering
\subfigure[Nhit distribution]{
\includegraphics[width=0.4\textwidth,angle=0]{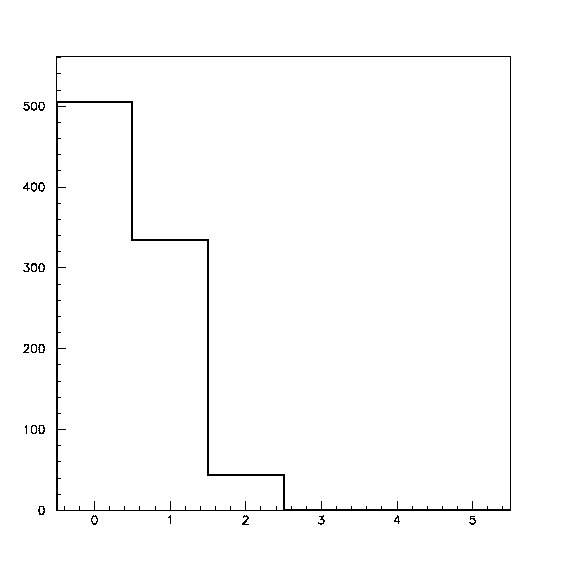}
}
\subfigure[Hits timing distribution]{
\includegraphics[width=0.4\textwidth,angle=0]{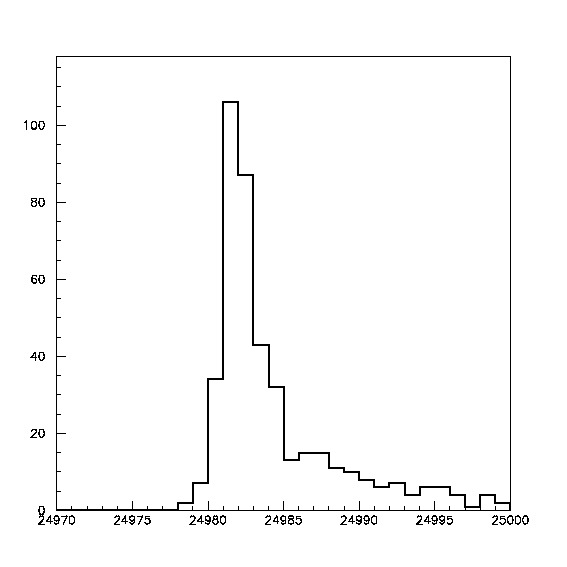}
}
\caption{\setlength{\baselineskip}{4mm}
(a) Number of hits in one trigger.
(b) Hit timing distribution. The horizontal axis is time counts, and 1 count corresponds to 2 ns.
Comparing with Fig.~\ref{Fig:wc_timing}, a
clear tail (slow component light) from diluted scintillator is present. 
} 
\label{FIG:SONLY_HIT}
\end{figure}

Figure~\ref{FIG:SONLY_HIT} shows the number of hits per one event and the hit timing 
distribution. Clearly, the slow components from the scintillator can be seen. 

\subsubsection*{\setlength{\baselineskip}{4mm}
Setting to detect scintillation plus Cherenkov light (Fig.~\ref{FIG:SETUPCS}(a))}
\indent

\begin{figure}[htbp]
\centering
\subfigure[Nhit distribution]{
\includegraphics[width=0.4\textwidth,angle=0]{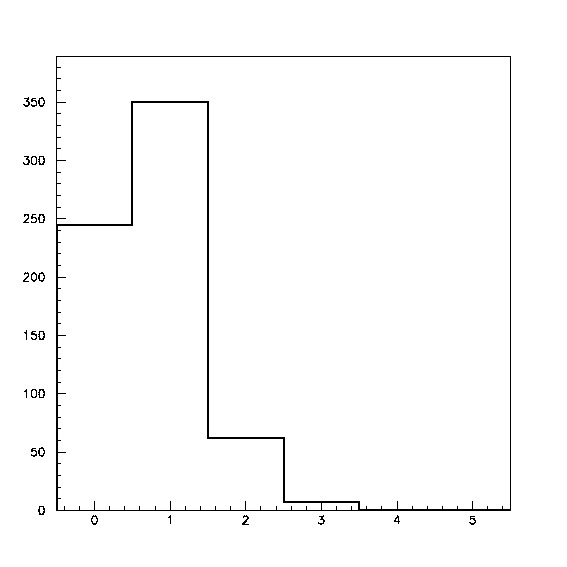}
}
\subfigure[Timing distribution]{
\includegraphics[width=0.4\textwidth,angle=0]{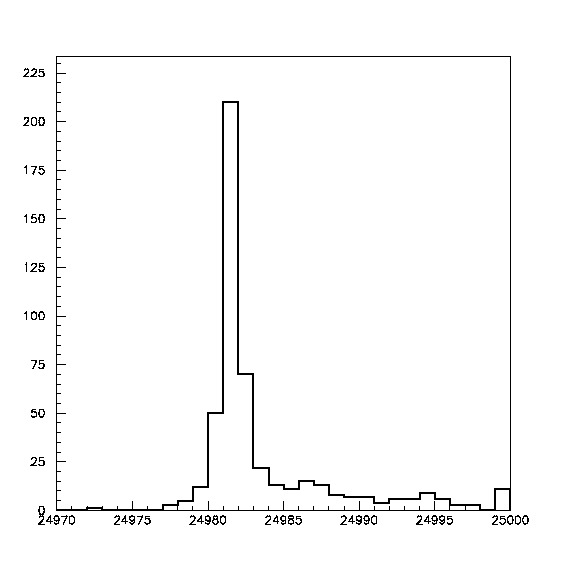}
}
\caption{\setlength{\baselineskip}{4mm}
(a) Number of hits in one trigger.
(b) Timing distribution of the hits (for Cherenkov+scintillation).
The horizontal axis is time counts, and 1 count corresponds to 2 ns.
} 
\label{FIG:SC_HIT}
\end{figure}

An analysis with the setting shown in 
Fig.~\ref{FIG:SETUPCS}(a) was performed as well.
The procedure for this analysis is exactly the same as what was 
described before.
Figure~\ref{FIG:SC_HIT} shows the results on the number of hits and 
timing of the hits. As expected, both the number of hits (left) and the
number of hits around the fast timing (right plot - same timing as the expected
Cherenkov timing), increased.

\subsubsection*{\setlength{\baselineskip}{4mm}
Comparison between scintillation only and Cherenkov+scintillation light}
\indent

\begin{figure}[htpb]
 \centering
 \includegraphics[width=0.55 \textwidth]{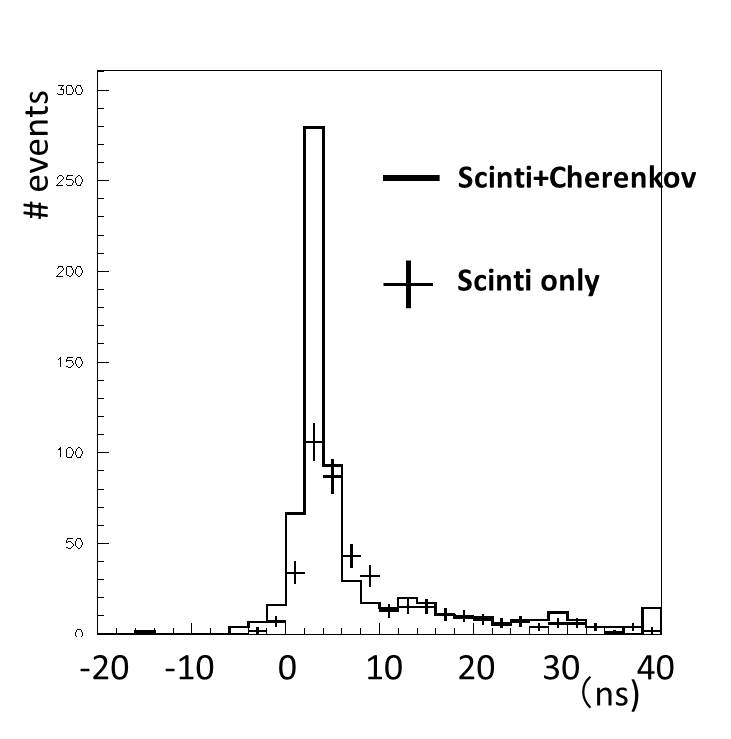}
\caption{\setlength{\baselineskip}{4mm} Timing comparison between scintillator only (cross) and Cherenkov + scintillator (histogram). Cherenkov light is concentrated in the fast timing, and the yield of Cherenkov is similar to the scintillation light. The horizontal axis is changed to \textit{ns} in order to improve understanding (starting point is still arbitrary).
}
 \label{Fig:timing_comp}
\end{figure}

Figure~\ref{Fig:timing_comp} shows the timing comparison between the scintillation 
light only case and the Cherenkov+scintillation light case. The horizontal 
axis is changed to \textit{ns} to understand it better (starting point is 
still arbitrary). Also both data sets are normalized using the number of triggers.
Two things are easily understood: (A) Cherenkov timing is faster 
than scintillation light as expected. (B) The amount of the Cherenkov light is 
comparable to the scintillation light in this scintillator condition. 
For the future work for item (A),
we have to compare this result with Fig.~\ref{Fig:timing} quantitatively 
(with smearing 1.8 ns at least), since one may reduce total jitter.
For the future work of item (B), to compare the amount of the light
from Cherenkov and scintillation, a detailed MC simulation 
is necessary, especially for the Cherenkov part since cosmic 
rays have momentum and zenith angle distributions.

\subsubsection{PID Summary}
\indent

We checked the performance of a LSND type liquid scintillator directly.
The Cherenkov light is useful information in this diluted scintillator 
condition as expected. However, this statement is still qualitative, and
we need a more quantitative statement in combination with the MC simulation.

Three future directions should be examined with this R$\&$D.
(1) variation of the concentration of the scintillator components for 
diluted scintillator, (2) similar study
using linear alkylbenzene (LAB), and (3) test beam with protons.
All will be done in order with priorities. 

With these inputs, it is important to estimate the realistic PID capability  
using MC simulation with a real size detector. This will be done within about 1.5 years. 

\subsection{Pulse Shape Discrimination (PSD) of Liquid Scintillator}
\subsubsection{R\&D of Gd loaded liquid scintillator(GdLS)}
\indent

The Japanese group tried to load Gd to LAB based liquid scintillator following the Daya-Bay experiment's paper~\cite{CITE:GdLS}. The JSNS$^2$ experiment will use this liquid type and the
test with a small size detector will be done in Japan.
In order to load Gd in LAB, it is necessary to use a Gd-carboxylate complex formed by a mixture of gadolinium chloride, neutralization solution with TMHA, and ammonium hydroxide. A sample of the Gd-carboxylate complex was made and used to produce the Daya Bay type liquid scintillator with 0.1 w\% of Gd concentration.

In order to check whether Gd was successfully loaded in the LAB, the thermal neutron capture time was measured since there is a correlation between the neutron capture time and Gd concentration.

\begin{figure}[htpb]
\begin{center}
\includegraphics[width=10cm]{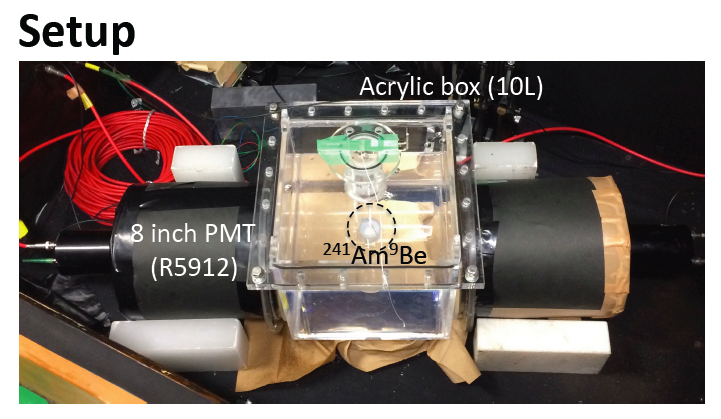}
\caption{\setlength{\baselineskip}{4mm}
Picture of the setup for measuring the thermal neutron capture time with the Daya Bay type GdLS.}
\label{FIG:GdLS}
\end{center}
\end{figure}

The measurement setup is shown in Fig.~\ref{FIG:GdLS}. 

\begin{figure}[htpb]
\begin{center}
\includegraphics[width=10cm]{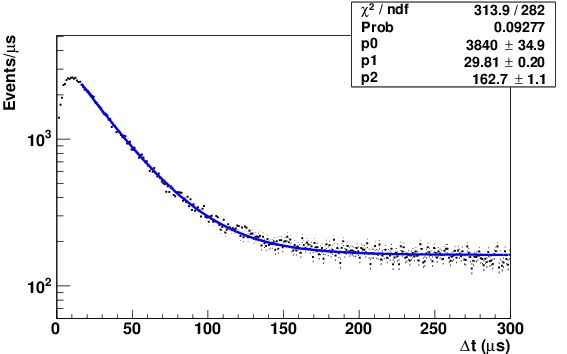}
\caption{\setlength{\baselineskip}{4mm}
Distribution of thermal neutron capture time using the Daya Bay type GdLS. The blue line shows the fit result with a sum of exponential and flat functions.}
\label{FIG:dt}
\end{center}
\end{figure}

The volume of the sample is 10L contained in an acrylic box. The scintillation light is viewed by two 8 inches PMTs and the signals are taken with a flash ADC module (CAEN V1721, 500MS/s, 8bits). An $^{241}$Am$^{9}$Be source, which emits both a 4.4 MeV gamma and neutron below 11 MeV, is set at the center of the acrylic box. The gamma and protons recoiled by the neutron are detected as the prompt signal, and gammas from thermal neutron capture on Gd are detected as delayed signal after about 30$\mu$sec of mean time.
Figure~\ref{FIG:dt} shows the time difference ($\Delta$t) distribution  between the prompt and delayed signals, which includes samples of accidental coincidences. Therefore, it was fit with a sum of an exponential and flat function. Then the thermal neutron capture time is estimated as 29.8$\pm$0.2 $\mu$sec, which can be compared with the expected time of 28 $\mu$sec. Considering the fit result, it was confirmed that about 0.09 w\% of the Gd concentration was loaded in LAB.

\subsubsection{R\&D of GdLS with capability of n/$\gamma$ pulse shape discrimination (PSD)}
\subsubsection*{Measurement of the PSD capability for Daya Bay type of GdLS}
\indent

We measured the PSD capability for Daya Bay type scintillator(DBLS) without Gd (only LAB, PPO and bisMSB).
The neutrino energy in JSNS$^2$ is a few tens MeV. The PSD capability increases with the neutrino energy, so if the DBLS shows enough cosmic induced fast neutrons rejection power while maintaining high signal efficiency in the energy range around a few tens MeV, we can use DBLS without any PSD capability improvement. 

\begin{figure}[htpb]
\begin{center}
\includegraphics[width=10cm]{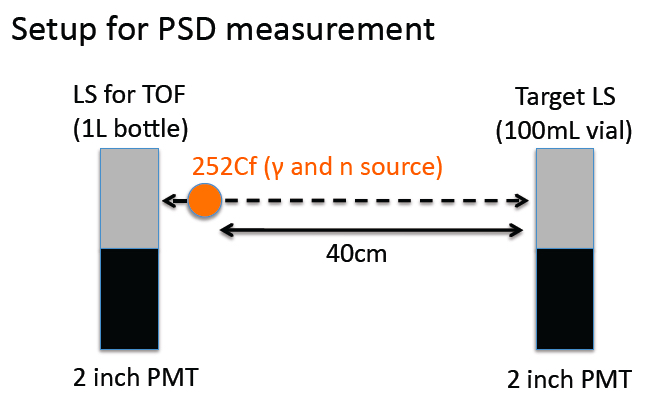}
\caption{Scheme of the PSD capability measurement.}
\label{FIG:PSDSETUP}
\end{center}
\end{figure}

Figure.\ref{FIG:PSDSETUP} shows the setup of the 
PSD capability measurement. 
The target scintillator is contained in a 100 mL vial and another liquid scintillator volume, with 1 L, where the $^{252}$Cf source is attached. The $^{252}$Cf source emits both $\gamma$ and neutron with an energy of few MeV, and the coincidence signals of the scintillators are detected for the PID. The distance between the target scintillator and the $^{252}$Cf source is 40cm. Using the hit time difference between both scintillators, the neutron events can be distinguished from the gamma events, because the time of flight(TOF) of the neutron is different from that of gammas, as shown in the vertical axis of Fig.~\ref{FIG:TOFPSD}.

Waveforms of  both scintillators are taken with a flash ADC module (CAEN V1730, 500MS/s, 14bits).
To evaluate the PSD capability, the ratio of the tail integrated charge of the waveform to the total integrated charge is used (TailQ/TotalQ variable). 
In this analysis, events around 500 photoelectrons are used for the evaluation
of PSD capability since this number of photoelectrons corresponds to the mean
energy of neutrons / $\gamma$s of the $^{252}$Cf source with consideration of the
light acceptance of the 2 inches PMT attached to the vial vessel. 
On the other hand, this number of photoelectrons corresponds to 3.5 MeV in
the JSNS$^2$ experiment considering the acceptance difference of the detectors.
 Neutrino energy is a few tens MeV in the JSNS$^2$ experiment, so this analysis gives a very conservative evaluation. To evaluate the PSD capability in the actual energy range, higher energy neutron and electron sources are necessary.

\begin{figure}[htpb]
\begin{center}
\includegraphics[width=14cm]{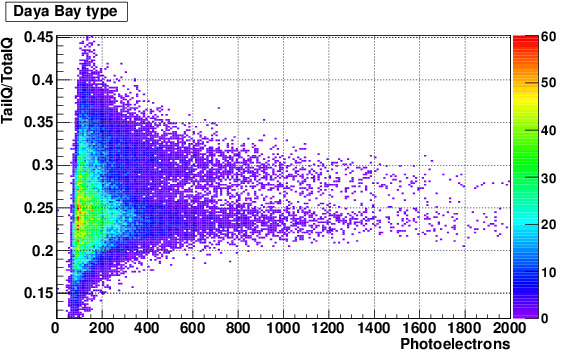}
\caption{\setlength{\baselineskip}{4mm}
  Correlation between the TailQ/TotalQ variable and the detected number of photoelectrons for the DBLS.
}
\label{FIG:PEPSD}
\end{center}
\end{figure}

\begin{figure}[htpb]
\begin{center}
\includegraphics[width=14cm]{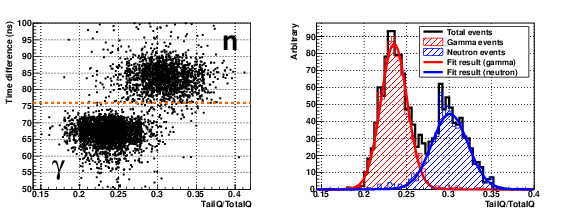}
\caption{\setlength{\baselineskip}{4mm}
A correlation between the TOF and the TailQ/TotalQ variable(left plot) around 500 photoelectrons in Fig.~\ref{FIG:PEPSD}, and 1D histograms of the TailQ/TotalQ  distributions for gammas and neutrons (right plot)}
\label{FIG:TOFPSD}
\end{center}
\end{figure}

\begin{figure}[htpb]
\begin{center}
\includegraphics[width=10cm]{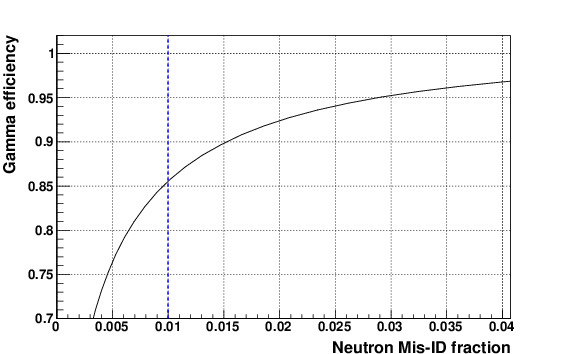}
\caption{\setlength{\baselineskip}{4mm}
Correlation between the detection efficiency of gammas and neutron Mis-ID fraction.}
\label{FIG:CutEff}
\end{center}
\end{figure}

 Figure~\ref{FIG:PEPSD} shows the correlation between the TailQ/TotalQ variable and the detected number of photoelectrons for the DBLS.
In the plot, there are two bands along the horizontal axis, which show the separation between gammas and neutrons. 
 The left plot of Fig.~\ref{FIG:TOFPSD} shows the correlation between the TOF and the TailQ/TotalQ variable around 500 photoelectrons in Fig.~\ref{FIG:PEPSD}, and the right plot shows the 1D histograms of the TailQ/TotalQ  distributions for gammas and neutrons. 
Using the TOF cut (orange line in the left plot), the neutron events can be distinguished from the gamma events. Each TailQ/TotalQ distribution after applying the TOF cut was fit with a Gaussian function, and the rejection power of neutrons and the detection efficiency of neutrino events (gamma) can be estimated by using the fit results. Figure~\ref{FIG:CutEff} shows the correlation between the detection efficiency of gammas and the neutron mis-ID fraction. 
At a factor of 100 rejection power for neutrons (neutron Mis-ID fraction $\sim$ 0.01), the detection efficiency of gammas for the DBLS is 0.85. The high detection efficiency using the DBLS is shown around 3.5 MeV equivalent already, so an efficiency increase is expected in the energy range of a few tens MeV due to photo-statistics.

\subsubsection*{Future plan}
\indent

It is important to measure the PSD capability in the actual JSNS$^2$ energy range of a few tens MeV. Neutrons and electrons created by an accelerator can be considered for this test. 

In addition to testing with a small size detector, it is essential to estimate 
the PSD capability of the real size detector using MC simulations.
Because of the distortion of the hit time distribution (or sum of waveforms) 
due to scintillation light attenuation, the light reflection by the tank and 
mis-reconstruction of the vertex, a large detector has worse PSD capability.
 
We need to design a detector which can precisely measure the time 
distribution of scintillation light emission event by event for high 
PSD capability. We are considering the following crucial items;\\
(A) Develop a tank with less light reflection\\
(B) Flash ADC module with high sampling rate($>$500MS/s) and number of bits($>$10bits).\\
(C) Optimization of number of photoelectrons (light yield of the GdLS). \\
(D) Develop a good vertex reconstruction algorithm.

\subsection{APD (SiPM)}
\indent

 The current design of the JSNS$^2$ detector is based on technologies that have succeeded in other experiments. The geometric size of the JSNS$^2$ detector, however, will be comparatively reduced to adapt to the MLF facility. With respect to the optical detection, large photomultipliers (10 inch) will be used as photon-detectors. In this case, large-sized PMT could affect the design of a sub-detector, such as a Veto counter. This influence becomes large in the compact detector system, such as with JSNS$^2$. Therefore, we have started to study the possibility of using APD (avalanche photodiode) technology as a substitute of the photomultiplier.  It is one possibility as a supplement to the JSNS$^2$ detector design.

 For the JSNS$^2$ test, Hamamatsu photonics provided some engineering samples of MPPC (multi-pixel photon counter), APD production. We have started the verification of the MPPC in liquid scintillator for long-term usage. At present, three types of MPPC were tested (listed in Table. \ref{APDlists}).

\begin{table}[htb]
 \begin{center}
 \caption{APD list}
  \begin{tabular}{|l|c|r|c|} \hline
	Model			&	Package Type	& Pixel	& Window material \\ \hline \hline
	S-10362-11-025U 	&	Metal		& 1600	& Borosilicate glass \\ \hline
	S-10362-33-050U	&	Ceramic		& 400	& Silicon \\ \hline
	S-10362-11-100U	&	SMD			& 100	& Epoxy resin \\ \hline
  \end{tabular}
  \label{APDlists}
   \end{center}
\end{table}

 In the JSNS$^2$ experiment, liquid scintillator(LAB) is chemically active. A MPPC may therefore be damaged by the liquid scintillator. We are studying potential damage to the window material due to the package type.

  For the first test, we used the mineral oil in section~\ref{SEC:SC}, and continue to observe an influence on the window of MPPC in liquid scintillator. The silicon window of the ceramic package was deformed and become cloudy in a day.
 
 In the future, we will evaluate the influence with a term of a month or year. We will check the deformation of the window and gain of the MPPCs. We have a plan for development of a MPPC in the liquid scintillator in collaboration with the manufacturer.

\section{\setlength{\baselineskip}{4mm} Background Measurement on the Proton Bunch Timing using 1.6L detector}  
\indent

As shown in the previous PAC~\cite{CITE:NOVSR}, there is a 
significant neutral particle flux that is observed in the proton 
bunch timing window, while this background is manageable
outside of this window for the JSNS$^2$. 
An additional particle ID measurement (neutrons or $\gamma$s) for the
particles on the proton bunch timing using a small size 
(1.6L) liquid scintillator 
detector was planned at the MLF 3rd floor. 
We assumed that all particles are neutrons in the previous status report, 
but, if there is some $\gamma$
fraction, the number of backgrounds for the delayed signal will be smaller.

Unfortunately, there have been two large incidents (fire from muon site and 
mercury target cooling issue) at the MLF. Therefore, the measurement was not
done yet, but the safety discussion was successfully done carefully between
facility people and the experiment. 
Hopefully, this measurement will be done starting this autumn.

\section{Summary and Plan}  
\indent

In this report, we show the status of our R$\&$D work, which
is going well. However, the concrete detector performance with
quantitative numbers with the realistic MC simulation 
should be shown before the stage-2 approval.

In addition to this R$\&$D work, a realistic detector location will be
investigated in consideration of the the facility constraints.

\section{Acknowledgements}
\indent

We warmly thank the MLF people, especially, Dr. Futakawa, MLF Division leader, 
the neutron source group, muon group and the user facility group for the 
various kinds of supports.
We also appreciate the support from grant-in-aid, J-PARC and KEK. 


\end{document}